\documentclass[apj]{emulateapj-rtx4}
\usepackage{apjfonts}
\usepackage{graphicx}
\usepackage{amssymb}
\usepackage{amsbsy}
\usepackage{amsmath}
\usepackage{mathrsfs}
\usepackage{color}

\DeclareMathAlphabet{\mathbi}{OT1}{ptm}{bx}{it}
\SetMathAlphabet\mathbi{bold}{OT1}{ptm}{bx}{it}

\def\hb{{\rm H\beta}}
\def\Oiii{{[O~{\footnotesize III}]}}

\slugcomment{To appear in the Astrophysical Journal Letters}
\shorttitle{A BAYESIAN METHOD FOR THE INTERCALIBRATION OF SPECTRA IN
RM}
\shortauthors{LI ET AL.}

\begin{document}
\title{A Bayesian Method for the Intercalibration of Spectra in
Reverberation Mapping}
\author{
  Yan-Rong Li\altaffilmark{1},  
  Jian-Min Wang\altaffilmark{1,2}, 
  Chen Hu\altaffilmark{1},
  Pu Du\altaffilmark{1},
  and Jin-Ming Bai\altaffilmark{3, 4}}
\affil{$^1$ Laboratory for Particle Astrophysics, Institute of High 
Energy Physics, Chinese Academy of Sciences, \\19B Yuquan Road, 
Beijing 100049, China; \href{mailto:liyanrong@mail.ihep.ac.cn}{liyanrong@mail.ihep.ac.cn}\\
$^2$ National Astronomical Observatories of China, Chinese 
Academy of Sciences, 20A Datun Road, Beijing 100020, China\\ 
$^3$ Yunnan Observatories, 
Chinese Academy of Sciences, Kunming 650011, China\\
$^4$ Key Laboratory for the Structure and Evolution of Celestial Objects, 
Chinese Academy of Sciences, Kunming 650011, China
}

\begin{abstract}
Flux calibration of spectra in reverberation mapping (RM) is most often
performed by assuming the flux constancy of some specified narrow emission lines,
which stem from an extended region that is sometimes partially spatially resolved, in 
contrast to the point-like broad-line region and the central continuum source. 
The inhomogeneous aperture geometries used among different observation sets in a joint monitoring
campaign introduce systematic deviations to the fluxes of broad lines and central continuum, 
and intercalibration over these data sets is required.
As an improvement to the previous empirical correction performed by comparing the (nearly)
contemporaneous observation points, we describe a feasible Bayesian method
that obviates the need for (nearly) contemporaneous observations, naturally incorporates 
physical models of flux variations, and fully takes into account the measurement errors. 
In particular, it fits all the data sets simultaneously regardless of samplings and
makes use of all of the information in the data sets. A Markov Chain Monte Carlo implementation 
is employed to recover the parameters and uncertainties for intercalibration. 
Application to the RM data sets of NGC 5548 with 
joint monitoring shows the high fidelity of our method.
\end{abstract}


\section{Introduction}
Reverberation mapping (RM) is a well-established technique for the study of 
broad-line regions (BLRs) in active galactic nuclei (AGNs) with broad emission 
lines (\citealt{Blandford1982, Peterson1993}). With appropriate analysis, RM experiments 
divulge the geometry, kinematic, and ionization structure information of BLRs  
(e.g., \citealt{Brewer2011, Pancoast2012, Li2013}). Over the 
past two  decades, the BLR size derived using the time delay between the 
continuum variation and the broad emission line response has been utilized 
with great success to measure the mass of the central supermassive black hole 
by combining it with the width of the broad emission line (e.g., \citealt{Peterson2004}). 
The tight relationship between BLR sizes and optical luminosities
of AGNs plays a key role in the demography of supermassive black holes
in large AGN surveys (e.g., \citealt{Bentz2013}, and references therein).

At present there are $\sim$50 nearby Seyfert galaxies and quasars with RM measurements in
the literature (e.g., \citealt{Bentz2013}), although a huge amount of effort has 
been invested in RM experiments. In practice, an RM campaign is quite observationally 
intensive and requires monitoring an object over a sufficient period with reasonable 
temporal resolution. Such high demand of time interval and sampling leads RM programs 
to be commonly undertaken by cooperative observations at multiple observatories, 
such as the well-known AGN Watch Project (\citealt{Peterson2002}) and MDM campaigns 
(\citealt{Denney2010, Grier2012}). Spectra calibration is most often based on the 
assumption that \Oiii~$\lambda$5007 line remains constant in flux over the 
timescale of interest and all spectra are scaled to an adopted absolute flux of  
\Oiii~$\lambda$5007, which can be measured on photometric nights
(\citealt{Peterson1991, vanGroningen1992}). The problem that arises with using \Oiii~$\lambda$5007
for such a calibration is that its emission region (narrow-line region; NLR) is sometimes
spatially resolved and the size is most likely comparable with or even larger 
than the aperture size, in contrast to the effectively point-like BLR and central 
continuum sources. Consequently, the inhomogeneous aperture geometries used among 
different observation sets in a joint monitoring campaign admit different
amounts of light from the NLR, and therefore introduce systematic deviations 
to the fluxes of broad emission lines (\citealt{Peterson1995}). Similarly, this effect 
also influences the central continuum fluxes contaminated by the host galaxy starlight. 

\cite{Peterson1995} proposed an empirical correction to such an aperture effect
by adopting one of the data set as standard, and applying a multiplicative scale factor 
and an additive flux adjustment to the other sets to bring the closely spaced measurements 
from the two sets into agreement. In reality, it is always impractical to base 
the correction on exactly contemporaneous observations. One has to relax such strict simultaneity
and instead use pairs of observations that are closely separated (usually by more than one days), 
depending on the sampling of each data set. This apparently degrades the highest achievable 
temporal resolution.

In this {\em Letter}, we describe a novel method for intercalibration of reverberation 
mapping data that obviates the need for (nearly) contemporaneous observations, naturally 
incorporates physical models of flux variations, and fully takes into account 
the measurement errors. The method is based on Bayesian statistics and is sufficiently 
elastic to automated program manipulation.

\section{Description of the Method}
\subsection{Variability Modeling}
While variability across all bandpasses is one of the outstanding characteristics 
of AGNs (\citealt{Ulrich1997}), its underlying mechanisms remain inconclusive.
There are recent extensive studies on the nature of AGN variability that are devoted to, 
but not limited to, exploring the ensemble properties of variability and 
its correlations with physical parameters of the AGN (e.g., \citealt{Zuo2012, Ai2013, 
MacLeod2012, Meusinger2013}); constructing the analytic stochastic description (e.g., 
\citealt{Kelly2009, Kelly2011, Kozlowski2010, MacLeod2010, Zu2013}); 
and testing/applying physically motivated models for variability (e.g., 
\citealt{Reynolds2009,Dexter2011, Pechacek2013} and references therein). Numerical 
simulations show that AGN variability is plausibly linked to some hydrodynamic or 
magnetohydrodynamic instabilities/turbulence within accretion disks, although 
such simulations are still in an early stage (e.g., \citealt{Noble2009, Reynolds2009}).
Using well sampled optical light curves of AGNs, it has been found that the optical 
power spectral density (PSD) of AGN variability can be described by a power law
$P(f)\propto f^{-\alpha}$ with flattening to a constant below some break frequency that 
typically corresponds to a timescale of dozens of days (\citealt{Czerny1999, Kelly2009, 
Kelly2011, MacLeod2010}). The slope $\alpha$ of the optical PSD seems to depend on the temporal 
frequency under consideration, changing from $\sim2$ on a timescale of days 
(\citealt{Kelly2009, Zu2013}), the normal temporal resolution of ground-based RM campaigns, 
to steeper values on much a shorter timescale, which, however, is based on a 
very preliminary analysis of the  {\em Kepler} data archive (\citealt{Mushotzky2011}). 
These features motivate the statistical modeling of variability of AGN optical continuum
by a damped random walk (DRW) process with great success (\citealt{Kelly2009, Kelly2011}).
This is further reinforced by subsequent investigations of large samples of AGN light curves 
(\citealt{Kozlowski2010, MacLeod2010, Zu2013, Li2013}).

Specifically, a DRW process is a stationary process such that its 
covariance function at any two times $t_1$ and $t_2$ depends only on the time difference
$t=|t_1-t_2|$ and has the form of
\begin{equation}
S(t) = \sigma^2\exp\left(-\frac{t}{\tau}\right),
\label{eqn_cov}
\end{equation}
where $\tau$ is the damping timescale of the process to return to its mean 
and $\sigma$ is the standard deviation of variation on long timescales ($\gg\tau$).
The corresponding PSD is a Lorentzian centered  at zero
\begin{equation}
P(f)=\frac{4\sigma^2\tau}{1+(2\pi\tau f)^2}.
\label{eqn_psd}
\end{equation}

Since the optical variability of AGNs is well described by a DRW process, it is 
expected that the variations of broad emission lines also follow DRW processes but with 
separate sets of $\tau$ and $\sigma$, according to the principle of RM that broad 
emission line variations are blurred echoes of the continuum variation 
(\citealt{Blandford1982, Peterson1993}). Apparently, there should be some correlations 
between the parameter sets of DRW processes for the optical continuum and the broad 
emission lines. 

It worth stressing that the present method is not only restricted to the DRW process. 
For more general cases, we make use of the fact that the covariance function and the 
PSD are Fourier duals of each other
\begin{equation}
P(f)= \int S(t)e^{-2\pi i f t}dt {~~~\text{and}~~~}
S(t) = \int P(f) e^{2\pi i f t} df.
\end{equation}
For any given AGN variability modeled by either a PSD or covariance function,
we can thereby construct the Bayesian posterior and perform intercalibration of the data 
sets as in next section.

\subsection{Intercalibration}
The measured data at hand are the flux time series of the continuum and broad emission lines
that have been calibrated with the specified narrow emission line (\citealt{vanGroningen1992}). 
Intercalibration is required to correct the effect of inhomogeneous apertures.
For illustration purposes, we adopt the flux light curves of the broad H$\beta$ and {5100\AA} 
continuum calibrated by \Oiii~$\lambda5007$. As proposed by \cite{Peterson1995}, 
after defining one data set for the target flux scale, 
the broad $\hb$ fluxes of the other data sets are corrected with respect to the target as 
\begin{equation}
F(\hb)=\varphi\cdot F(\hb)_{\rm obs},
\label{eqn_fhb}
\end{equation}
and the continuum 5100 {\AA} fluxes as
\begin{equation}
F_{\lambda}(5100\text{\AA})=\varphi\cdot F_\lambda(5100\text{\AA})_{\rm obs}-G,
\label{eqn_fcon}
\end{equation}
where the subscript ``obs'' means the measured values, $\varphi$ is a scalar for 
point-source correction, and $G$ is a flux offset for extended source correction 
(e.g., host galaxy starlight). The values of $\varphi$ and $G$ depend on the 
individual data set. It is apparent that $\varphi=1$ and $G=0$ for the target set.
We note that \cite{Zu2011} proposed using different means for the time series of 
each data set, which are quite trivial to obtain, to reconcile the different levels 
of the host galaxy contamination. Here, the parameter $G$ is mathematically equivalent 
to their proposal in effect, but obviates extra steps for calculating 
the differences of the means required for intercalibration.

We now develop a Bayesian framework to perform intercalibration 
with the variation modeling described in the preceding section.
We first derive the likelihood probability for the continuum fluxes. 
Let the column vector $\mathbi{y}_{\rm c}$ 
denote the ``intrinsic'' continuum fluxes  and $\mathbi{f}_{\rm c}$ denote 
the $m$ corresponding measurements subjected to aperture effect
in a joint-monitoring campaign with $k$ data sets. 
The intrinsic light curve is deemed to be the sum of 
an underlying variation signal $\mathbi{s}_{\rm c}$ described by
a DRW process and a constant $q_{\rm c}$ 
representing the mean of the light curve  (\citealt{Zu2011, Li2013}), i.e.,
\begin{equation}
\mathbi{y}_{\rm c}=\mathbi{s}_{\rm c}+\mathbi{E}q_{\rm c},
\end{equation}
where $\mathbi{E}$ is a vector with all unity elements.
From Equation~(\ref{eqn_fcon})
and taking into account the measurement errors $\mathbi{n}_{\rm c}$, 
$\mathbi{f}_{\rm c}$ is generated by $\mathbi{y}_{\rm c}$
\begin{equation}
\mathbi{f}_{\rm c}=\boldsymbol{\Phi}^{-1}(\mathbi{y}_{\rm c}+\mathbi{LG}) 
+ \mathbi{n}_{\rm c},
\end{equation}
where $\boldsymbol{\Phi}$ is an $m\times m$ diagonal matrix
whose diagonal elements are formed out of the $k$ multiplicative factors $\varphi$ for 
$k$ data sets, $\mathbi{G}$ is a vector of the $k$ additive factors $G$, and 
$\mathbi{L}$ is an $m\times k$ matrix with entries of $(0,...,0, 1, 0, ..., 0)$ for 
$i$th data set.

As usual, we assume that both $\mathbi{s}_{\rm c}$ and 
$\mathbi{n}_{\rm c}$ are Gaussian and uncorrelated.
By following similar procedures in \cite{Zu2011} and \cite{Li2013}
based on the framework outlined by 
\cite{Rybicki1992}, we can trivially obtain the likelihood probability for
$\mathbi{f}_{\rm c}$ as%
{\footnote{Note that there is a typo in the normalization factor of
Equations (3), (5), and (A7) of \cite{Li2013}. 
The correct form is given in Equation (17) of \cite{Zu2011}.}}
\begin{eqnarray}
&&P(\mathbi{f}_{\rm c}|\boldsymbol{\Theta}, \sigma_{\rm c}, \tau_{\rm c})\nonumber\\
&=&\int P(\mathbi{s}_{\rm c})P(\mathbi{n}_{\rm c})P(q_{\rm c})\nonumber\\
&\times&
\delta\left[\mathbi{f}_{\rm c}-\boldsymbol{\Phi}^{-1}(\mathbi{y}_{\rm c}+\mathbi{LG}) 
- \mathbi{n}_{\rm c}\right]d^m\mathbi{s}_{\rm c}d^m\mathbi{n}_{\rm c}dq_{\rm c}\nonumber\\
&=&
\frac{\sqrt{|\boldsymbol{\Phi}^T\mathbi{N}_{\rm c}\boldsymbol{\Phi}|}}
{\sqrt{(2\pi)^{m-1}|\mathbi{C}_{\rm c}||\mathbi{N}_{\rm c}||\mathbi{E}^T\mathbi{C}_{\rm c}^{-1}\mathbi{E}|}}\nonumber\\
&\times& 
\exp\left[-\frac{1}{2}\left(\hat{\mathbi{y}}_{\rm c}-\mathbi{E}\hat q_{\rm c}\right)^T\mathbi{C}_{\rm c}^{-1}
\left(\hat{\mathbi{y}}_{\rm c}-\mathbi{E}\hat q_{\rm c}\right)\right],
\label{eqn_pcd}
\end{eqnarray}
where $\delta[x]$ is the Dirac function, the superscript ``$T$'' represents the transposition, 
$\mathbi{C}_{\rm c}\equiv \mathbi{S}_{\rm c} + \boldsymbol{\Phi}^T\mathbi{N}_{\rm c}\boldsymbol{\Phi}$,
$\mathbi{S}_{\rm c}$ is the covariance matrix of the signal $\mathbi{s}_{\rm c}$ given by
Equation (\ref{eqn_cov}), and $\mathbi{N}_{\rm c}$ is the covariance matrix 
of the noise $\mathbi{n}_{\rm c}$,
\begin{equation}
\hat{\mathbi{y}}_{\rm c} = \boldsymbol{\Phi}\mathbi{f}_{\rm c} - \mathbi{LG}~~~~~~{\rm and}~~~~~~
\hat q_{\rm c} = \frac{\mathbi{E}^{T}\mathbi{C}_{\rm c}^{-1}\hat{\mathbi{y}}_{\rm c}}
{\mathbi{E}^{T}\mathbi{C}_{\rm c}^{-1}\mathbi{E}},
\end{equation}
where $\hat q_{\rm c}$ is indeed the best estimate of $q_{\rm c}$.
Here the integral over $q_{\rm c}$ marginalizes it and its prior probability $P(q_{\rm c})$
is assumed to be constant. 
In Equation (\ref{eqn_pcd}), the free parameters to be determined 
are the $k-1$ sets of $(\varphi, G)$ for intercalibration,
denoted by $\boldsymbol{\Theta}$, and two parameters ($\sigma_{\rm c}, \tau_{\rm c}$) 
for the variation modeling of the continuum. 

In a similar way, we can readily write out the likelihood probability 
$P_{l}(\mathbi{f}_l|\boldsymbol{\Theta}, \sigma_{l}, \tau_{l}) $
of the emission line $\mathbi{y}_{l}$
by replacing the subscript ``c'' with ``$l$'' in above equations.
For the sake of simplicity, we treat the measurements
of the continuum $\mathbi{f}_{\rm c}$ and emission line $\mathbi{f}_{l}$
separately and assume that $\mathbi{f}_{\rm c}$ and $\mathbi{f}_{l}$ are 
independent. As such, we can obtain a simple form of the posterior probability 
according to the Bayes' theorem:
\begin{eqnarray}
&&P(\boldsymbol{\Theta}, \sigma_{\rm c}, \tau_{\rm c}, \sigma_{l}, \tau_{l}|\mathbi{f}_{\rm c},
\mathbi{f}_{l})\nonumber\\
&&\quad\propto
P(\boldsymbol{\Theta}, \sigma_{\rm c}, \tau_{\rm c}, \sigma_{l}, \tau_{l})
P_{\rm c}(\mathbi{f}_{\rm c}|\boldsymbol{\Theta}, \sigma_{\rm c}, \tau_{\rm c}) 
P_{l}(\mathbi{f}_{l}|\boldsymbol{\Theta}, \sigma_{l}, \tau_{l}),
\label{eqn_post}
\end{eqnarray}
where $P(\boldsymbol{\Theta}, \sigma_{\rm c}, \tau_{\rm c}, \sigma_{l}, \tau_{l})$
is the prior probability of the free parameters.
Maximizing Equation (\ref{eqn_post}) yields the best estimate 
of the free parameters. The uncertainties of the
free parameters are determined from a Markov Chain Monte Carlo (MCMC) analysis and
added in quadrature to flux intercalibration. 
Equations (\ref{eqn_fhb}) and (\ref{eqn_fcon})
are then applied for the final intercalibrated fluxes.
We point out that since the posterior distribution is constructed over 
all the data sets, we can select any data set as the standard one, 
provided its measurements are sufficiently good to represent the genuine fluxes.

Compared with the previous empirical method proposed by \cite{Peterson1995}, 
the superiority of the present Bayesian method lies at (1) performing intercalibration
on all the data sets simultaneously in Equation (\ref{eqn_post}) and therefore can 
make use of all the information
in the data sets, (2) obviating the need for simultaneity of the data sets 
and relaxing the requirements for the sampling rates,
(3) not degrading the highest achievable temporal 
resolution of the campaign, (4) capability of naturally incorporating physical models 
of variations and taking into account the measurement errors, and 
(5) sufficiently feasible for automated program manipulation regardless of samplings.

\subsection{Markov Chain Monte Carlo Implementation}
We employ an MCMC analysis to explore the statistical properties of the free parameters.
The samples of the free parameters are constructed from Markov chains for the posterior 
probability distribution (see Equation (\ref{eqn_post}))
using parallel tempering and the Metropolis-Hastings algorithm
(\citealt{Liu2001}). Here, the parallel tempering algorithm guards the Markov chain from
being stuck in a local maximum and facilitates its convergence to globally optimized 
solutions. The prior probabilities in Equation (\ref{eqn_post}) are assigned as follows:
for parameters whose typical values ranges are known, a uniform prior is assigned; otherwise,
if the parameter information is completely unknown, a logarithmic prior is assigned.
Among the free parameters, the priors for ($\sigma_{\rm c}, \tau_{\rm c}$), 
($\sigma_{l}, \tau_{l}$), and $k-1$ free parameters $\varphi$ 
are set to be logarithmic, and the rest are set to be uniform.
The Markov chain is run for 150,000 steps in total. 
The best estimates for the parameters are taken to be the expectation value
of their distribution and the uncertainties are taken to be the standard deviation.

\section{Tests And Application}
To demonstrate the fidelity of our new method, we apply it to the publicly 
accessible RM database of the Seyfert galaxy NGC 5548 ($z=0.0167$) from the AGN Watch 
Project (\citealt{Peterson2002}). NGC 5548 was jointly monitored for as long as 13 yr 
between 1989 and 2001 by numerous ground-based optical telescopes, making it well suited 
for verifying our method. Also, the original observed values of $F_{\lambda}(5100~\text{\AA})$ 
and $F(\text{H}\beta)$ (i.e., without calibration)
were published in the series of papers for NGC 5548 (\citealt{Peterson2002}, and references therein).
This allows us to directly compare the intercalibration results from our new method with those from
the previous empirical method. We make use of the RM data from the first 2 yr (1989 and 1990) 
tabulated in \citet{Peterson1991} and \citet{Peterson1992}, respectively.
The absolute flux of \Oiii~$\lambda5007$ for NGC 5548
is set to $F(\text{[\rm O{\footnotesize~III}]}\lambda5007)=5.58\times10^{-13}\rm erg~s^{-1}$,
as determined by \cite{Peterson1991}. The observation noise values are assumed to be uncorrelated 
so that the covariance matrix $\mathbi{N}$ in Equation (\ref{eqn_pcd}) is diagonal.

\begin{figure*}[th!]
\centering
\includegraphics[angle=-90.0, width=0.89\textwidth]{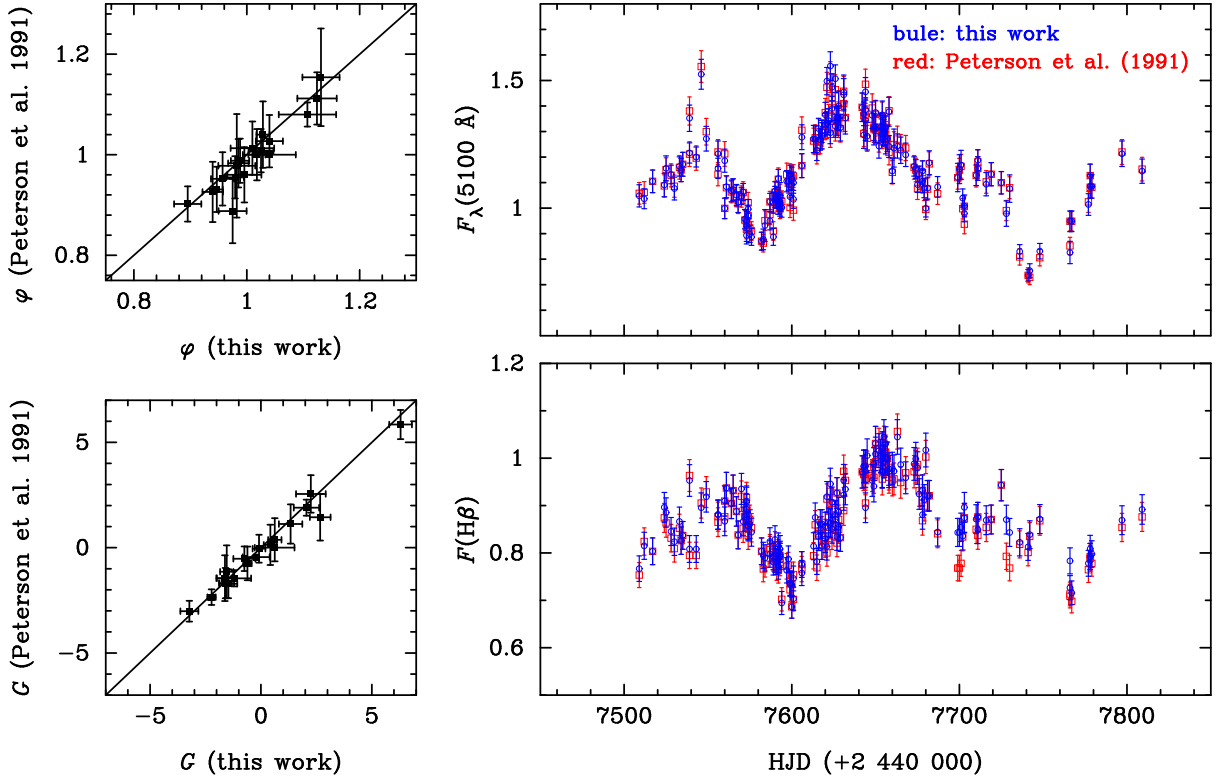}
\caption{Intercalibration of the RM data of NGC 5548 from the year 1989 extracted from 
\cite{Peterson1991}. Left panels: comparison of the intercalibration factors $\varphi$ (top)
and $G$ (bottom) derived in this work with those from
\cite{Peterson1991}. The unit of $G$ is $10^{-15}~{\rm erg~s^{-1}~cm^{-2}~\text{\AA}^{-1}}$.
Solid lines represent the equality of the two results.
Right panels: intercalibrated light curves of the 5100 {\AA} continuum (top) and
broad H$\beta$ emission line (bottom). The continuum  and H$\beta$ fluxes are in units of 
$10^{-14}~{\rm erg~s^{-1}~cm^{-2}~\text{\AA}^{-1}}$ and $10^{-12}~{\rm erg~s^{-1}~cm^{-2}}$,
respectively. The light curves from \cite{Peterson1991} are superposed for comparison.}
\label{fig_year1991}
\end{figure*}
\begin{figure*}[th!]
\centering
\includegraphics[angle=-90.0, width=0.89\textwidth]{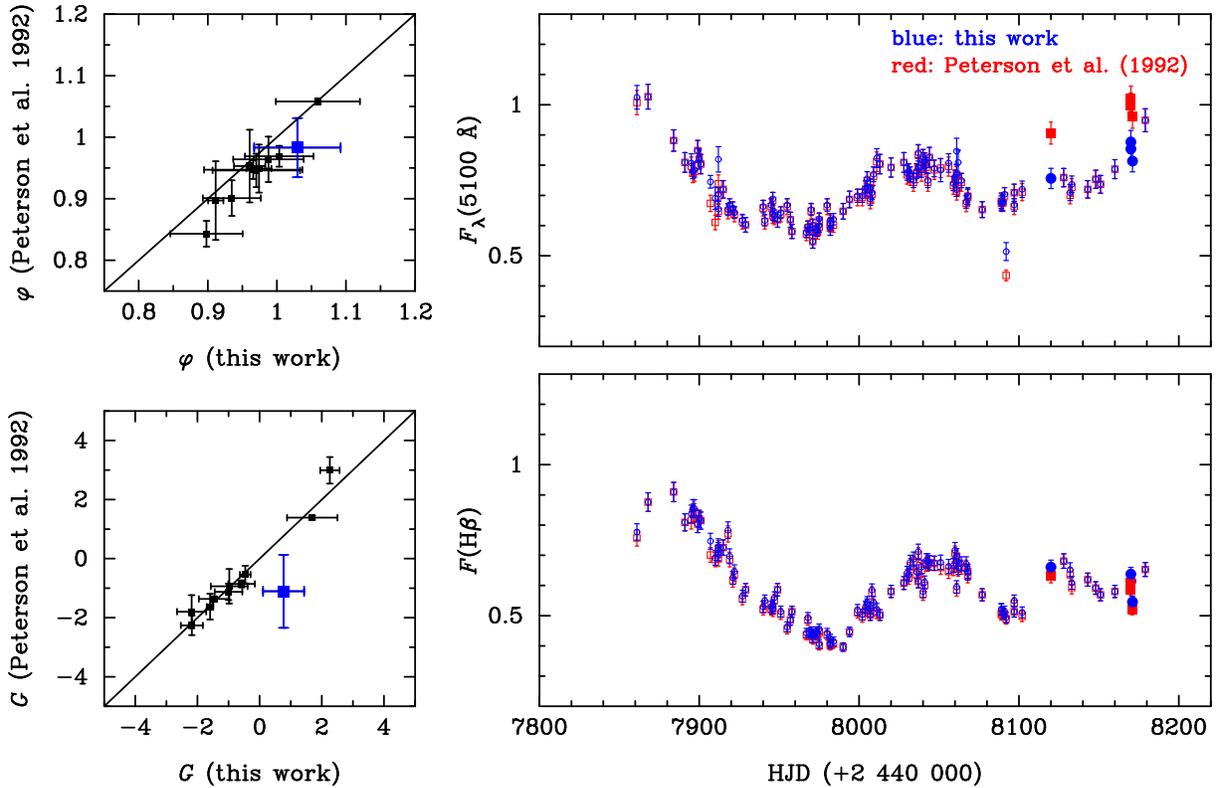}
\caption{Same as Figure \ref{fig_year1991} but for the RM data from the year 1990 extracted from 
\cite{Peterson1992}.  Squares highlighted in blue in the left panels represent the intercalibration 
constants for set ``K'', where the previous method encountered difficulties obtaining 
reliable intercalibration in \cite{Peterson1992}.
Solid points in the right panels highlight the intercalibrated fluxes for set ``K'' in this 
work (in blue) and in \cite{Peterson1992} (in red).}
\label{fig_year1992}
\end{figure*}

\subsection{Year 1989}
In the year 1989, NGC 5548 was monitored by 17 telescopes with various
individual apertures. \cite{Peterson1991} grouped these measurements into 20 
data sets according to the aperture sizes of the corresponding instruments 
(see Tables 6-8 therein). For the sake of clarity, in this work, we adopt the 
data set codes following \cite{Peterson1991} to identify the instruments that 
obtained the spectrum. 
Each data set is regarded to be internally homogeneous and the variations due to 
seeing are subsumed into the measurement uncertainties. \cite{Peterson1991} selected the 
data set (identified by code ``A'') with fairly numerous observations as the reference to 
gain a reasonable overlap with other data sets. They found that to improve the accuracy of 
the intercalibration, they needed to compare all the measurement 
pairs separated by up to 2 days. Therefore, the temporal resolution of their light curves 
should be degraded to be at least larger than 2 days. Indeed, the average interval between 
the measurements is 3.1 days and the median interval is 1 day (see \citealt{Peterson1991}
for details).

In Figure~\ref{fig_year1991}, we compare the intercalibration results from our method 
with those from \cite{Peterson1991}. The fractional measurement uncertainties for the 
continuum and the H$\beta$ fluxes are set uniformly to be 0.040 and 0.035, respectively 
(\citealt{Peterson1991}). We find remarkable agreement
of our results with those of \cite{Peterson1991}, indicating the fidelity of our method. 
As mentioned above, our method does not degrade the temporal resolution; however, 
the homogeneous {\em IUE} data of NGC 5548 showed evidence that the variability seems 
to begin to appear on a timescale longer than $\sim$5 days (\citealt{Peterson1991}). 
It is no surprise that the light curves of both the continuum and H$\beta$ fluxes are 
fairly consistent with \cite{Peterson1991}'s results. Nevertheless, the measurement 
point of the continuum on HJD 2,447,546 that seems anomalously high in \cite{Peterson1991} 
is now slightly lower in our results, as plotted in the top right panel of Figure~\ref{fig_year1991}.

\subsection{Year 1990}
In the year 1990, NGC 5548 was monitored by 12 telescopes and all the measurements are grouped into
12 data sets by \cite{Peterson1992} (see Tables 5-7 therein). Again, data set ``A'' was adopted 
as the reference. The short time scale sampling in this year is
not sufficiently good as the year in 1989. Consequently, to obtain reliable intercalibration accuracy, 
\cite{Peterson1992} adopted different time separations of the measurement pairs for comparison
with different data sets. In particular, based on their method,
it is impossible to intercalibrate data set ``K'' because none of its observations are 
within 5 days of any observations of the other sets. Since 
the variation of NGC 5548 begins to be notable on a timescale of $\sim5$ days,  
\cite{Peterson1992} used the intercalibration constants for the set ``K'' in 
\cite{Peterson1991}. However, an inspection of the intercalibration constants 
obtained in \cite{Peterson1991, Peterson1992} clearly shows that the constants 
are not exactly identical for the same set code. Moreover, while the resulting fluxes of H$\beta$ of 
set ``K'' looks quite good, the fluxes of the continuum, highlighted with solid symbols 
in Figure~\ref{fig_year1992}, are unexpectedly high (the measurement HJD 2,448,120 in particular), 
implying that the intercalibration for set ``K'' is plausibly doubtful. 

We show our intercalibration results for the year 1990 in Figure~\ref{fig_year1992}. Again, except for 
set ``K'', there is quite good agreement between the two methods. However, our intercalibrated fluxes 
of the continuum for set ``K'' seems much more reasonable and are consistent with the other 
closely spaced measurements regarding the variation trend. This is because our Bayesian approach 
is not based on a comparison of the closely spaced measurements but instead looks for the most optimized
solution for maximizing the posterior distribution in Equation (\ref{eqn_post}). This permits 
us to cope with poorly sampled data as in set ``K''. 

\section{Conclusions}
We propose a feasible Bayesian method for spectral intercalibration in
a joint monitoring campaign based on the assumption of flux constancy of 
some specified narrow emission line (e.g., \Oiii~$\lambda5007$). Compared with 
the previous empirical method comparing the closely spaced measurement pairs,
our new method obviates the requirement for (nearly) contemporaneity of observations
and takes into account the measurement errors naturally. The Bayesian approach enables 
us to perform intercalibration on all the data sets simultaneously and self-consistently 
regardless of sampling rates, and therefore can cope with poorly sampled data.
Application to the RM database of NGC 5548
from the AGN Watch Project shows the fidelity of our method and its capability to 
yield appropriate intercalibration where the previous method encountered difficulties.

In conclusion we propose a road map for complete spectral calibration in RM campaigns in which one 
or more emission lines with constant flux are present: 
first employ the algorithm described by \cite{vanGroningen1992}
to perform relative scaling based on the adopted emission line, which takes into account the
zero-point wavelength-calibration errors between individual spectra and resolution differences;
and then employ our method to perform intercalibration to correct for the effect
of inhomogeneous apertures.

%===================================================================================
\acknowledgements{We thank the referee for constructive comments that significantly 
improve the manuscript.
This research is supported by NSFC-11133006, 
11173023, 11233003, and 11303026, the China-Israel NSFC-ISF 11361140347,
and the Strategic Priority Research Program -
The Emergence of Cosmological Structures of the Chinese Academy of Sciences, grant No. XDB09000000.}
%===================================================================================

\end{document}